\documentstyle[multicol,aps,epsfig]{revtex}
\begin{document}
\newcommand{\be}{\begin{equation}}
\newcommand{\ee}{\end{equation}}
\newcommand{\lb}{\label}
\newcommand{\en}{\epsilon}
\newcommand{\ven}{\varepsilon}
\newcommand{\bF}{{\bf f}}
\newcommand{\bu}{{\bf u}}
\newcommand{\bv}{{\bf v}}
\newcommand{\br}{{\bf r}}
\newcommand{\bk}{{\bf k}}
\newcommand{\bx}{{\bf x}}
\newcommand{\bD}{{\bf D}}
\newcommand{\bK}{{\bf K}}
\newcommand{\bS}{{\bf S}}
\newcommand{\om}{\omega}
\newcommand{\vl}{\overline{{\bf u}}}
\newcommand{\vs}{{\bf u}^{\prime}}
\newcommand{\oll}{\overline{\omega}}
\newcommand{\os}{{\omega}^{\prime}}
\newcommand{\ul}{\overline{{\bf u}}}
\newcommand{\us}{{\bf u}^{\prime}}
\newcommand{\pll}{\overline{p}}
\newcommand{\el}{\overline{e}}
\newcommand{\oL}{\overline}
\newcommand{\btau}{{\mbox{\boldmath $\tau$}}}
\newcommand{\bdot}{{\mbox{\boldmath $\cdot$}}}
\newcommand{\bdots}{{\mbox{\boldmath $:$}}}
\newcommand{\btimes}{{\mbox{\boldmath $\times$}}}
\newcommand{\grad}{{\mbox{\boldmath $\nabla$}}}
\newcommand{\bsigma}{{\mbox{\boldmath $\sigma$}}}
\newcommand{\bomega}{{\mbox{\boldmath $\omega$}}}
\newcommand{\bll}{{\mbox{\boldmath $\ell$}}}

\topmargin-0.5in
\pagestyle{myheadings}
\draft

\title{Is the Kelvin Theorem Valid for High-Reynolds-Number Turbulence?}

\author{Shiyi Chen$^{1-4}$, Gregory L.
Eyink$^{1,2,3}$,\\
Minping Wan$^{1}$, and Zuoli Xiao$^{1}$}
\address{
${}^{1}$Department of Mechanical Engineering,
${}^{2}$Applied Mathematics \& Statistics, \\
The Johns Hopkins University,
Baltimore, MD 21218\\
${}^{3}$Center for Nonlinear Studies and T-Division,
Los Alamos National Laboratory, Los Alamos, NM 87545\\
${}^{4}$College of Engineering and CCSE, Peking
University, China\\
}
\maketitle
\begin{abstract}
The Kelvin-Helmholtz theorem on conservation of circulations is supposed to
hold
for ideal inviscid fluids and is believed to be play a crucial role in
turbulent
phenomena, such as production of dissipation by vortex line-stretching.
However,
this expectation does not take into account singularities in turbulent velocity
fields at infinite Reynolds number.  We present evidence from numerical
simulations
for the breakdown of the classical Kelvin theorem in the three-dimensional
turbulent energy
cascade. Although violated in individual realizations, we find that
circulations are still
conserved in some average sense. For comparison, we show that Kelvin's theorem
holds for individual realizations in the two-dimensional enstrophy cascade, in
agreement
with theory. The turbulent  ``cascade of circulations''  is shown to be a
classical analogue
of phase-slip due to quantized vortices in superfluids and various applications
in geophysics
and astrophysics are outlined.
\end{abstract}


\begin{multicols}{2}
\narrowtext

The theorem on conservation of circulations due to Helmholtz
\cite{Helmholtz1858} and Kelvin \cite{Kelvin1869} is a fundamental
fluid dynamical result with many important consequences. The circulation
invariants are a topological  obstruction to ideal vortex-reconnection in
classical and quantum fluids \cite{KidaTakaoka94,Barenghietal01}.
Ertel's theorem \cite{Ertel42} on conservation of potential vorticity in
geophysical
fluid dynamics is a differential version of the Kelvin theorem
\cite{Pedlosky98}.
The conservation of circulations was also argued by G. I. Taylor
\cite{Taylor17,TaylorGreen37,Taylor38} to play a key  role in the enhanced
production
of dissipation in turbulent fluids, by the process of vortex line-stretching.
More recently, regularizations of the Navier-Stokes equation have been proposed
as model equations for large-scale turbulence \cite{Holmetal98,Foiasetal01},
motivated by requiring that a Kelvin theorem be preserved. However, despite
the frequent application to turbulent flows, existing proofs of conservation
of circulation are valid only for smooth, laminar solutions of the ideal fluid
equations. One might naively expect that the conservation will hold better as
the viscosity decreases, or the Reynolds number increases, but this has never
been shown. This Letter presents evidence from numerical simulations both
for the breakdown of the classical Kelvin theorem and for a generalized version
that is still valid in turbulent flow at infinite Reynolds number.

We consider the incompressible Navier-Stokes equation in space dimension
$d\geq 2:$
\be \partial_t \bu + (\bu\bdot\grad)\bu = -\grad p + \nu\bigtriangleup \bu,
       \,\,\,\,\,\,\,\,\,\, \grad\bdot\bu=0, \lb{INS} \ee
where $\bu(\bx,t)$ is the velocity, $p(\bx,t)$ the kinematic pressure (or
enthalpy),
and $\nu$ the kinematic viscosity. The classical Kelvin theorem in this context
states
that for any closed, rectifiable loop $C$ at an initial time $t_0,$ the
circulation
$\Gamma(C,t)=\oint_{C(t)}\bu(t)\bdot d\bx$ satisfies
\be {{d}\over{dt}}\Gamma(C,t)=\nu\oint_{C(t)}\bigtriangleup\bu(t)\bdot d\bx,
\lb{kelvin} \ee
where $C(t)$ is the loop advected by the fluid velocity, at time $t.$
In the inviscid limit $\nu\rightarrow 0,$ it follows heuristically from
equation
(\ref{kelvin}) that the circulation is conserved for any initial loop $C.$
This conservation law can be shown to be a consequence of  Noether's
theorem for an infinite dimensional gauge symmetry group of the ideal
fluid model associated to particle-relabelling \cite{Arnold66,Salmon88}.

The conservation can be anomalous, however, in the zero-viscosity limit,
if the velocity field becomes too singular. In that case, the righthand side
of (\ref{kelvin}) need not vanish as $\nu\rightarrow 0.$ The situation is quite
similar to that for conservation of energy, which, as observed by Onsager
\cite{Onsager49}, can also become anomalous in the sense that the
energy dissipation $\varepsilon=\nu|\grad\bu|^2$ need not vanish in
the inviscid limit. For the energy integral, the existence of a dissipative
anomaly can be associated to a turbulent cascade with a constant
mean flux of energy to arbitrarily small length-scales. See
\cite{Eyink94,Constantinetal94,DuchonRobert00}.

Anomalous circulation conservation may be formulated in a similar fashion
\cite{Eyink06}. Let $\oL{\bu}_\ell=G_\ell*\bu$ denote the low-pass
filtered velocity at length-scale $\ell,$ where
$G_\ell(\br)=\ell^{-d}G(\br/\ell)$
is a smooth filter kernel. Then $\oL{\bu}_\ell$ satisfies the following
equation
(dropping  viscous terms that are small in the inertial-range):
\be \partial_t\oL{\bu}_\ell+(\oL{\bu}_\ell\bdot \grad)\oL{\bu}_\ell
   = -\grad\oL{p}_\ell+\bF_\ell
   , \lb{F-euler} \ee
where $\oL{p}_\ell$ is the large-scale (modified) pressure and where
$\bF_\ell$ is the {\it turbulent vortex-force}:
\be \bF_\ell=\oL{(\bu\btimes \bomega)}_\ell
                    -\oL{\bu}_\ell\btimes\oL{\bomega}_\ell. \lb{v-force} \ee
Let us choose a rectifiable closed loop $C$ in space. We define
 $\oL{C}_\ell(t)$ as the loop $C$ advected by the filtered velocity
 $\oL{\bu}_\ell$ and  a ``large-scale circulation'' with initial loop $C$
 as the line-integral
$ \oL{\Gamma}_\ell(C,t) = \oint_{\oL{C}_\ell(t)}
                               \oL{\bu}_\ell(t)\cdot d\bx. $
It follows from (\ref{F-euler}) that
\be
(d/dt)\oL{\Gamma}_\ell(C,t) = \oint_{\oL{C}_\ell(t)}\bF_\ell(t)\cdot d\bx
\lb{F-kelvin} \ee
The righthand side of (\ref{F-kelvin}) represents a turbulent transport
of lines of large-scale vorticity $\oL{\bomega}_\ell(t)$ out of the loop
$\oL{C}_\ell(t)$. Using formula (\ref{v-force}) and the righthand rule,
it is easy to see that $\bF_\ell$ provides a torque that reduces the
large-scale circulation around the loop when a vortex-line migrates
outside it.  This process can be seen to be local-in-scale, with most of the
transport due to scales close to $\ell$. Therefore, it is natural to think
of the process as a ``circulation-cascade" \cite{Eyink06}. This motivates
the definition, for any loop $C$ and filter length $\ell,$ of a
{\it circulation-flux}
%
$ K_\ell(C)= -\oint_{C} \bF_\ell  \cdot d\bx.  $
If the effect of the modes at lengths $<\ell$ is, on average,to diffuse
the lines of large-scale vorticity, then this definition implies that the
signs of $\oL{\Gamma}_\ell(C,t)$ and $K_\ell(C,t)$ should be positively
correlated.

The righthand side of (\ref{F-kelvin})---or the circulation-flux---does not
need to vanish in the limit taking first $\nu\rightarrow 0,$ then $
\ell\rightarrow 0$. This provides a purely inviscid mechanism for violation
of the Kelvin theorem, if the fluid velocity is sufficiently singular. A
simple estimate of the vortex force shows that $|\bF_\ell|=O(|\delta
\bu(\ell)|^2/\ell),$ where $\delta\bu(\ell)$ is the velocity increment
across a length $\ell$ \cite{Eyink06}. Suppose that the velocity is
H\"{o}lder continuous in space with exponent $h,$ so that $|\delta
\bu(\ell)|=O(\ell^h)$. It follows then that
$K_\ell(C)=O\left(\ell^{2h-1}\right)$
for any loop $C$ of finite length \cite{Eyink06}. Thus, the Kelvin theorem will
be valid under these assumptions if $h>1/2.$ The latter condition will hold in
certain cases, e.g. the $2d$ enstrophy cascade, where it is expected that $h=1$
(with logarithmic corrections).  However, in the $3d$ energy cascade, the
condition is not expected to be true, since even the mean-field Kolmogorov
exponent $h=1/3$ is $<1/2.$ Thus, the rigorous results do not settle the
issue of whether Kelvin's theorem is valid for infinite Reynolds-number
turbulence in $3d$, where a non-vanishing circulation-flux is possible.

To explore these questions, we have carried out a direct numerical
simulation (DNS) of eq. (\ref{INS}) on a $1024^3$ periodic grid using
a pseudo-spectral parallel code with full dealiasing and time-stepping
by a second-order Adam-Bashforth method. The kinetic energy
was forced in the first two shells \cite{CS} and a statistical
stationary state was achieved after 5 large-eddy turnover
times. The final Taylor-scale Reynolds number was $Re_\lambda=383$
and about a decade of inertial range was achieved with constant mean
energy flux and a spectral exponent close to $-5/3.$ We investigated
the statistics of the circulation and circulation-flux on square loops of
edge-length $R.$ The line-integrals were evaluated by using Stokes theorem
and then calculating the surface-integrals over the square faces. For each
value of $R$ all square loops were considered on the computational grid
with that edge-length, at all positions in the periodic domain and with all
three possible orientations.

In Fig.~1(a) plot the probability density function (PDF) of the
circulation-flux
$K_\ell(C)$ for a square loop with radius $R=64$ (lattice units) and for
several
filter lengths $\ell<R.$ As $\ell$ decreases through the inertial-range, the
PDF's
approach a form that is nearly independent of the filter length. This invariant
PDF is non-Gaussian with stretched exponential tails, qualitatively similar
to the PDF of the circulation itself \cite{Caoetal96}. The crucial point here
is lack of dependence of the PDF on the filter-length $\ell,$ which is
analogous to the independence of the mean energy flux in the inertial-range.
Note that the mean of the circulation-flux must be zero in homogeneous
turbulence because of the exact identity $\bF_\ell=-\grad\bdot\btau_\ell+
\grad \kappa_\ell$ for the turbulent vortex-force, where $\btau_\ell=
\oL{(\bu\,\bu)_\ell}-\oL{\bu}_\ell\oL{\bu}_\ell$ is the turbulent subscale
stress and $\kappa_\ell=(1/2){\rm tr}\,\btau_\ell$ is the subscale kinetic
energy. However, higher-order moments of $K_\ell(C)$ can be non-vanishing.
In Fig.1(b) we plot the rms value of the circulation-flux as a function of
$k_c=\pi/\ell$ for several values of $R$ in the inertial range. The near
independence of rms flux to $k_c$ verifies the existence of the circulation
cascade. In an infinitely long inertial-range, for $\nu\rightarrow 0,$ this
nonlinear
cascade would lead to breakdown of the classical Kelvin theorem. It is also
interesting to consider the $R$-dependence of the inertial-range value of
the rms circulation-flux. In the inset to Fig.1(b) we plot the rms value versus
$R,$ revealing a scaling very close to $R^{1/2}.$ Since $K_\ell(C)$ has the
dimension of velocity-squared, the mean-field Kolmogorov scaling would
be $R^{2/3}.$ The smaller exponent shows that rms circulation-flux scales
anomalously with $R,$ similar to the circulation itself \cite{Caoetal96}.

  \begin{figure}[hbt]

    \begin{center}
      \includegraphics[width=7.5cm]{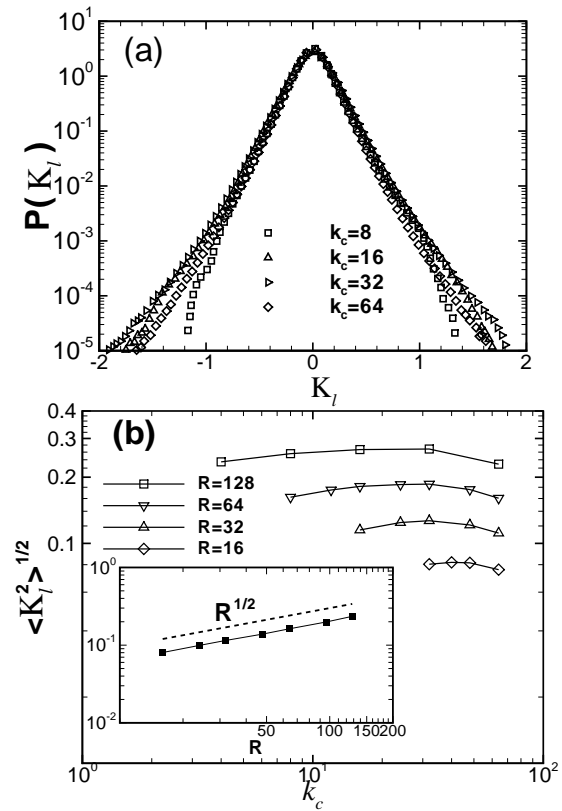}
    \end{center}

   \caption{(a) PDF of the circulation-flux for loops with radius
$R=64$ and for several filter lengths $\ell<R$.
(b) The rms value of the circulation-flux as a function of $k_c$
for various loop sizes $R$. The inset plots the plateau rms value
versus $R$.}

   \end{figure}

  It is illuminating to compare these numerical results for $3d$ with
 corresponding results for the enstrophy  cascade in $2d.$ We have
 analyzed the solutions of a $2048^2$ DNS, the details of which are
 given in \cite{Chenetal03}.  This simulation yielded about a decade
 and a half of inertial-range with constant mean enstrophy-flux and
 an energy spectral exponent close to $-3$. In Fig.2(a) we plot the PDF's
 of circulation-flux $K_\ell(C)$ from this DNS for a square loop with
 $R=128$ and several filter-lengths $\ell<R.$ Unlike the corresponding
 results in Fig.1(a) for $3d,$ the PDF's are not independent of $\ell$ but
 instead narrow rapidly as $\ell$ decreases. This is quantified in
 Fig.2(b), which plots the rms circulation-flux versus $k_c,$ again
 for loops with radius $R=128.$ A power-law decay is observed for increasing
$k_c$ with an exponent between $-1$ and $-2,$ consistent with the rigorous
 bound in \cite{Eyink06}. These results show that the Kelvin theorem is
 valid in the $2d$ enstrophy cascade, whereas circulations are not
 conserved in the $3d$ energy cascade range as $\ell\rightarrow 0,$ due
 to persistent nonlinear transport of vortex-lines.

  \begin{figure}[hbt]

    \begin{center}
      \includegraphics[width=7.5cm]{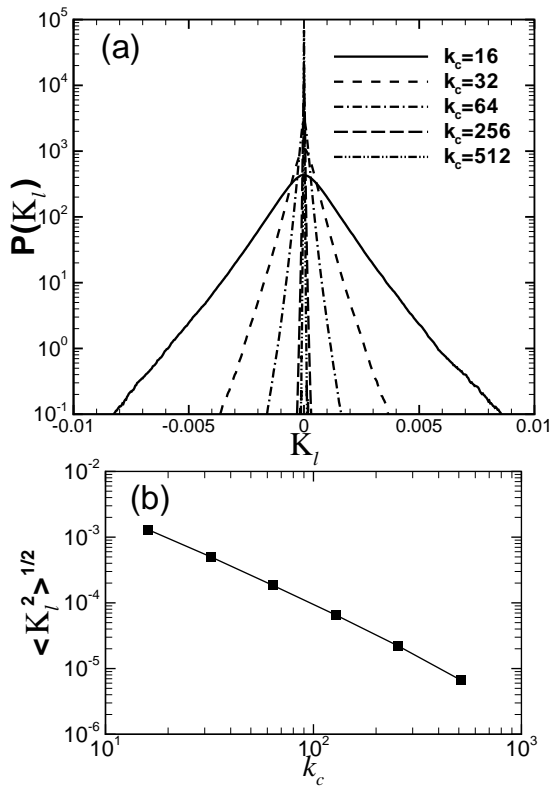}
    \end{center}

   \caption{(a) The PDF's of circulation-flux from a $2048^2$ simulation of
            2D enstrophy cascade for loops of radius $R=128$ and several
            $k_c=\pi/\ell$. (b) The rms circulation-flux plotted versus $k_c$,
            also for $R=128.$}

   \end{figure}

Although the Kelvin theorem is not valid in the classical sense in
$3d$ turbulence, it was conjectured in \cite{Eyink06} that a weaker
form may still be valid. Consistent with the results above, the
circulation $\Gamma(C,t)$ on an advected loop $C(t)$ is expected
not to be invariant in time. In fact, associated with the phenomenon
of  ``spontaneous stochasticity'' in the zero-viscosity limit
\cite{Bernardetal98,Chavesetal03}, the loop $C(t)$ should be
random for a {\it fixed} initial loop $C$ and advecting velocity
field $\bu.$  Under these circumstances, the time series of
$\Gamma(C,t)$ will be a stochastic process. It was proposed in
\cite{Eyink06} that the random time-series of circulations should
possess the ``martingale property'':
\be \langle \Gamma(C,t)|\Gamma(C,s),\,s<t' \rangle =
                     \Gamma(C,t'),\,\,\,\, t>t'. \lb{mart} \ee
That is, the conditional expectation of the circulation in the future
should be the last given value. This is a natural statistical
generalization of the Kelvin theorem and, in fact, is a generalized
form of the inviscid Euler equations of motion for a turbulent
fluid \cite{Eyink06}.

A full test of these ideas will be quite difficult and involve,
among other things, careful Lagrangian tracking of the loops $C(t).$
Here we check a somewhat weaker result. In Fig.3 we plot
the conditional average circulation flux $\langle K_\ell(C)|
\Gamma(C)=\Gamma\rangle$ from the $3d$ DNS as a function
of $k_c=\pi/\ell,$ for a square loop with $R=64$ and for various
values of the circulation level $\Gamma.$ Unlike (\ref{mart}), this
expectation includes an average over the turbulent ensemble of
velocities $\bu.$  The plot in Fig.3 shows that the sign of $K_\ell(C)$
is positively correlated with that of $\Gamma(C),$ consistent
with the diffusive character of turbulent vortex-line transport.
Thus, the effect of  subscale modes at length-scales $<\ell$
will be to reduce the magnitude of the circulation, regardless
of its sign. However, the conditional average flux for each
value of $\Gamma$ tends to zero as $\ell$ decreases through
the inertial range. This is a true effect of the nonlinear dynamics,
as illustrated by the dashed line in Fig.3, which plots the
expectation of the viscosity term $\nu\oint_{C}\bigtriangleup
\oL{\bu}_\ell \bdot d\bx$ conditioned upon $\Gamma=\Gamma_{rms},$
as a function of $k_c.$ Clearly viscous effects are negligible over
the range considered. These results show that the Kelvin theorem,
although violated in individual realizations, is still
valid for $3d$ turbulence in some average sense.

  \begin{figure}[hbt]

    \begin{center}
      \includegraphics[width=7.5cm]{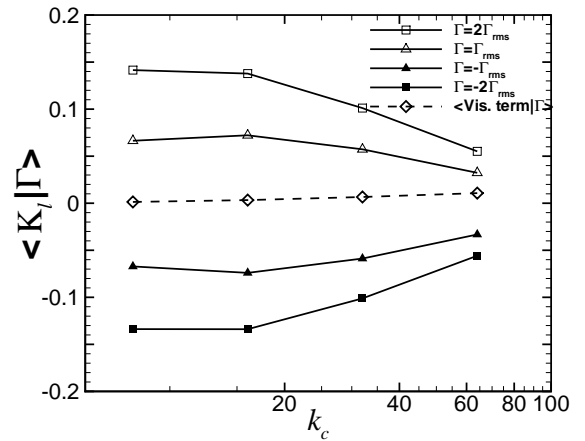}
    \end{center}

   \caption{The conditional average circulation-flux from
            the 3d DNS as a function of $k_c$, for a square loop
            with $R=64$ and for various values of the circulation level
            (solid lines). The dashed line shows the conditional expectation
            of the viscosity term, given $\Gamma=\Gamma_{rms},$ as a
            function of $k_c$.}

   \end{figure}

It is worth pointing out that the ``circulation cascade''
verified for fluid turbulence in this work is a classical
analogue of the ``phase slip'' phenomenon in superfluids,
such as ${\,\!}^4{\rm He}$ below the $\lambda$-point \cite{Anderson66}.
For example, consider the decay of a superfluid flow in a
thin toroidal ring \cite{Muelleretal98}. The decay is mediated
by the (thermal or quantum) nucleation of quantized vortices
which migrate out of the ring. The passage of a vortex across
the toroidal cross-section induces by phase-slip a pulse of torque
which decreases the circulation around the ring. It was emphasized
already by Anderson \cite{Anderson66} that phase-slip occurs
in superfluids only because the quantized vortices are not
material objects moving with the fluid. This is possible because
of the singular vortex core, where the superfluid density drops
to zero. The mechanism of turbulent circulation-cascade that
we have considered is quite similar. The vortex lines in a
turbulent flow are also not material, because singularities
in the velocity field allow them to diffuse relative to the fluid.
Unlike in superfluids, this is a continuous process, since
classical vortices are not quantized. There is also no need
for the singularities to be nucleated as fluctuations, since they
are everywhere present in the turbulent flow. Finally, we note
that in turbulence the diffusion of vortex-lines is not persistent in
scale, on average, and does not lead to irreversible decay
of circulations. As shown by Fig.~3, the partial effect of the
modes adjacent in scale $(\ell\sim R)$ leads to a mean decay
but, for $\ell\ll R,$ the circulation is conserved on average.
In this limit, time-reversal symmetry is restored.

The results in this work have many important implications.
The turbulent transport of vortex-lines, persistent as
$\ell\rightarrow 0,$ provides an inviscid mechanism for
vortex-reconnection and other changes of line topology.
However, a statistical form of the Kelvin theorem seems to
survive, which is crucial to justify Taylor's vortex-stretching mechanism
of turbulent energy  dissipation \cite{Taylor17,TaylorGreen37,Taylor38}.
Further research must clarify exactly to what extent conservation
of circulation survives for turbulent solutions of fluid equations.
Numerical studies must be extended to Lagrangian tracking of
advected loops $C(t).$ Theoretical investigations are required
to take into account the presumably fractal nature of such loops
\cite{SreenivasanMeneveau86}. Circulation-cascade should
occur also in superfluid turbulence \cite{Barenghietal01},
mediated in part by phase-slip of quantized vortex-lines.
Results similar to the present ones may be developed also for
conservation of magnetic flux (Alfv\'{e}n's theorem \cite{Alfven43}),
important in magnetohydrodynamic turbulence, and Ertel's theorem
\cite{Ertel42} in geophysical turbulence. The breakdown of
Alfv\'{e}n's theorem and the ``frozen-in'' character of magnetic
flux-lines is of particular importance  to account for the fast
reconnection observed in a variety of astrophysical turbulent
flows \cite{LazarianVishniac00}.

\vspace{.2in}

{\bf Acknowledgements.} We wish to thank R. E. Ecke, C. Meneveau,
K. R. Sreenivasan and E. T. Vishniac for useful conversations.  Simulations
were
run on the computer clusters supported by NSF grants \# CTS-0079674,
ASE-0428325,
at Johns Hopkins University and in the Center for Computational Science
and Engineering at Peking University, China.

\end{multicols}
\end{document}